
\overfullrule=0pt\magnification=\magstep1\baselineskip=20pt plus.1pt minus.1pt
\hsize=6.25 true in\vsize=9.65 true in\parindent=10 true pt
\font\reffont=cmcsc10\font\bigbf=cmbx10 scaled\magstep2
\font\sectfont=cmsl10 scaled\magstep2
\def\newline{\par\noindent}\def\newpage{\vfill\eject}
\def\startnumberingat#1{\pageno=#1\footline={\hss\tenrm\folio\hss}}
\newcount\eqnno\eqnno=0\newcount\secn\secn=0
\def\numeqn{\global\advance\eqnno by 1 \eqno(\the\eqnno)}
\def\rememeqn#1{\edef#1{\the\eqnno}}
\def\REF#1#2#3#4#5#6{\item{[#1]}{\reffont#2}{\it\ #3}{\bf\ #4}{\ (#5),\ #6}}

\newcount\refno\refno=0\def\ref#1{$\,^{[#1]}$}
\def\nextref#1{\global\advance\refno by 1\edef#1{\the\refno}}
\def\newsec#1\par{\vskip0pt plus .2\vsize\penalty-250
 \vskip0pt plus -.2\vsize\bigskip\vskip\parskip\global\advance\secn by 1
 \message {\the\secn. #1}\leftline{\sectfont\the\secn. #1}\nobreak\smallskip}
\def\beginsec#1\par{\vskip0pt plus .2\vsize\penalty-250
 \vskip0pt plus -.2\vsize\bigskip\vskip\parskip\message{#1}
 \leftline{\sectfont #1}\nobreak\smallskip}

\def\half{\textstyle{1\over 2}}\def\fourth{{1\over 4}}
\def\12{{1\over2}}\def\oa{{\bar a}}\def\ourchi{{\raise.5ex\hbox{$\chi$}}}
\def\csqw{\cos^2\Theta_W}
\def\bz{{\bar Z}}\def\gz{2\csqw\bz}
\pageno=0\nopagenumbers
\rightline{DAMTP 93--39}
\rightline{hep-ph/9308223}
\bigskip\bigskip
\centerline{\bigbf STABILITY OF TWO DOUBLET}
\centerline{\bigbf ELECTROWEAK STRINGS}
\vskip 1 true in
\centerline{MICHAEL A. EARNSHAW}
\centerline{MARGARET JAMES}
\bigskip
\centerline{\it Department of Applied Mathematics and Theoretical Physics}
\centerline{\it Silver Street, Cambridge, CB3 9EW, Great Britain}
\vskip 1 true in
\centerline{\hfill{\bf ABSTRACT}\hfill}\nobreak\smallskip\noindent

Vortex solutions in the two Higgs doublet electroweak model are constructed,
and
their stability to small perturbations is studied. The most general
perturbation
is decomposed into angular momentum modes, the least stable mode is identified,
and the linearised energy change of the vortex under this perturbation is
calculated numerically for various choices of parameters, thus determining
whether or not the string is stable. It is found that, for realistic values of
the Higgs mass and Weinberg angle, the string is unstable.

\newpage
\startnumberingat1
\newsec Introduction

In the last two years, interest has
been generated in the study of `embedded' defects. These occur when a theory
which has topological defects is contained in a larger theory and the defects
remain solutions of the field equations. General conditions for this to take
\nextref\vachaspatiA
place have been studied by Vachaspati et al\ref{\vachaspatiA}. A crucial
question is whether the defect is still stable. Generally we will not expect
them to remain topologically stable and the stability\footnote*{We are only
considering the stability under small perturbations. Hence, the word
`stability' should always be understood as `meta-stability'.} becomes a
dynamical question.

In this paper we consider embedded vortex solutions. A
\nextref\vachaspatiB
prototype example is the semi-local string\ref{\vachaspatiB}; the
Nielsen-Olesen
\nextref\nielsen
vortex\ref{\nielsen} is embedded in the semilocal model where
the gauge group $SU(2)_{global}\times U(1)_{local}$ is spontaneously broken to
$U(1)_{global}$ by a complex Higgs doublet. The vacuum manifold of the Higgs is
now $S^3$ and since the first homotopy group of this is trivial the solution is
no longer topologically stable. However the
solutions may still be classified into topological sectors separated by
infinite energy since finite energy requires that the Higgs field at spatial
infinity lies on a $U(1)$ orbit on $S^3$. The question is then whether the
solution is dynamically stable in a given sector. There is one free parameter
in the model, the ratio of the quartic coupling to the gauge coupling and the
string is stable for a finite
\nextref\hindmarsh\nextref\vachaspatiC
parameter range.\ref{\hindmarsh,\vachaspatiC}

When the $SU(2)$ symmetry is gauged we recover the electroweak model,
$SU(2)_L\times U(1)_Y\rightarrow U(1)_{em}$. The vortex solution persists in
\nextref\vachaspatiD\nextref\nambu
the model for any value of the Weinberg angle\ref{\vachaspatiD,\nambu}. Now all
configurations are in the vacuum sector i.e. any solution can be continuously
deformed to the vacuum. The parameter space is now two-dimensional, the
parameters being the ratio of the Higgs-mass to the Z-mass and the Weinberg
\nextref\vachaspatiE\nextref\james
angle. The string has been shown to be stable for a range in parameter
space\ref{\vachaspatiE,\james}.
However taking the experimental lower bound on the scalar Higgs and the
observed
value of the Weinberg angle, it was shown that if the minimal standard model is
the physically realised model then electroweak-strings are unstable. The
calculation was refined by including finite
\nextref\vachaspatiF
temperature and quantum effects\ref{\vachaspatiF} and while the stability of
the string is improved the physical values are still outside the region of
stability for temperatures close to the electroweak phase transition. This is
somewhat disappointing since the prospect of a stable classical vortex is an
exciting one. It could possibly be produced in future particle accelerators.
However it is not yet understood how to calculate the cross-section for what,
in quantum theory would be a coherent state of many particles and it may well
be that the available phase space is so small as to make the cross-section
negligible. In cosmology if the strings are meta-stable, there would be
implications for physics at the electroweak phase transition. Davis and
\nextref\davis
Brandenberger\ref{\davis} have suggested that they could play a role in
baryogenesis. We are therefore led to consider
extensions of the standard model where the stability region might be enlarged
sufficiently to include the physical values.
\nextref{\vachaspatiG}
Vachaspati and Watkins\ref{\vachaspatiG} suggested that the
vortex could be stabilised by the introduction of a further scalar field which
would form a bound state with the string.

A different possibility, the one
we investigate here, is that of extending the standard model via the
introduction of a second Higgs doublet. For simplicity we do not consider the
most general such model but restrict ourselves by imposing CP-invariance on the
Higgs potential. The model has seven parameters: the Weinberg angle, five Higgs
couplings and the ratio of the norm of the vacuum expectation values of the two
Higgs doublets. The stability analysis follows the same scheme as that in
reference [\vachaspatiE] although the introduction of the second Higgs doublet
somewhat complicates the analysis. In section 2. we discuss the vortex
solutions
in the two Higgs model. In section 3. we consider the perturbations around the
string solution and show that they decompose into channels labelled by angular
momentum and the eigenvalues of two discrete symmetry operators. In section 4.
we locate the least stable mode and reduce the analysis to solving a pair of
coupled Schr\"odinger equations. In section 5. we discuss the numerical work
and present our results and in 6. we give our conclusions.

\newsec Two Higgs-Doublet Electroweak Strings

We consider the minimal extension of the Higgs sector of the standard model by
introducing a second Higgs complex doublet. We give the most general potential
subject only to the restrictions of gauge invariance and that it has the
discrete
\nextref\higgs
symmetry,$(\Phi_1\leftrightarrow-\Phi_1,\Phi_2\leftrightarrow-\Phi_2)$
\ref{\higgs}. (The latter is required to ensure that flavour changing neutral
currents are not too large.)
$$\eqalign{V(\Phi_1,\Phi_2)=&\half
\lambda_1(\left\vert\Phi_1\right\vert^2-v_1^2)^2+
\half\lambda_2(\left\vert\Phi_2\right\vert^2-v_2^2)^2+\half\lambda_3
(\left\vert\Phi_1\right\vert^2 +\left\vert\Phi_2\right\vert-v_1^2-v_2^2)^2\cr
&+\lambda_4\left((\Phi_1^\dagger\Phi_1)(\Phi_2^\dagger\Phi_2)-\vert
\Phi_1^\dagger\Phi_2\vert^2\right)+\lambda_5\left ({\rm
Re}(\Phi_1^\dagger\Phi_2)-v_1v_2\cos\eta\right)^2\cr &+\lambda_6 \left({\rm
Im}(\Phi_1^\dagger\Phi_2)-v_1v_2\sin\eta\right)^2\cr} \numeqn$$
The vacuum expectation of the Higgs fields are given by
$$\langle\Phi_1\rangle=\pmatrix{0\cr v_1\cr},\ \ \langle\Phi_2\rangle=
\pmatrix{0\cr v_2e^{i\eta}\cr}.\numeqn$$
If $\sin\eta\neq 0$ then there is
CP-violation in the Higgs sector. This case is considerably more complicated
than the $\sin\eta=0$ case where, as we shall see later, the vortex solution is
CP-invariant and the perturbations then split into channels labelled CP-even
and CP-odd. Hence we set $\sin\eta=0$.
Also, the $\lambda_6$ term is the mass term for the CP-odd neutral scalar,
$A_0$, which will decouple from our analysis, so we are free to set
$\lambda_6=\lambda_5$. Then the last two terms are replaced by a single term
$$\lambda_5\left\vert(\Phi_1^\dagger\Phi_2)-v_1v_2 \right\vert^2.\numeqn$$

We are considering static, bosonic field configurations.
There is no time dependence and we choose a gauge where the zero components of
the gauge fields are set to zero. The energy functional is given by
$$E=\int d^3x\left[\fourth G_{ij}^a G_{ij}^a + \fourth F_{ij} F_{ij} +
(D_j\Phi_\oa)^{\dagger} (D_j\Phi_\oa)+V(\Phi_1,\Phi_2)\right]\numeqn$$
\rememeqn\eqnint
where $\oa=1,2$\footnote*{Barred indices always take the values 1 and 2.} and
$$G^a_{ij}=\partial_iW^a_j-\partial_jW^a_i+g\epsilon^{abc}W^b_iW^c_j,\numeqn$$
$$F_{ij}=\partial_iB_j-\partial_jB_i,\numeqn$$
$$D_i\Phi_{\oa} =\partial_i\Phi_{\oa}-{\textstyle{1\over2}}ig\tau^aW^a_i
\Phi_{\oa}-{\textstyle{1\over2}}ig'B_i\Phi_{\oa}\,\numeqn$$
where we take $T^a=-{1\over 2}i\tau^a$ as our basis for $L(SU(2))$ (denoting
the Lie algebra of $SU(2)$) with $\lbrack
T^a,T^b\rbrack =\epsilon^{abc}T^c$. The Weinberg angle is given by tan
$\Theta_W={g'/g}$. The semiclassical masses of the W-boson and the Z-boson
are, respectively,
$$M_W^2=\half g^2(v_1^2+v_2^2),\qquad
M_Z^2=\half\alpha^2(v_1^2+v_2^2),\numeqn$$
where $\alpha=(g'^2+g^2)^{1\over 2}.$

There are five physical Higgs bosons, a charged pair $H^{\pm}$, two
CP-even scalars, $H^0$ and $h^0$, and the CP-odd neutral scalar $A_0$ which we
have already discussed. The charged pair have mass
$m_{H_{\pm}}^2=\lambda_4(v_1^2+v_2^2)$
and the two remaining Higgs scalars mix through the mass matrix,
$$M=\pmatrix{2v_1^2(\lambda_1+\lambda_3)+v_2^2\lambda_5&(2\lambda_3+\lambda_5)
v_1v_2\cr(2\lambda_3+\lambda_5)
v_1v_2&2v_2^2(\lambda_2+\lambda_3)+v_1^2\lambda_5\cr}\numeqn$$
\rememeqn\eqnmas
and the physical eigenstates have masses
$$m_{H_0,h_0}^2=\half\left(M_{11}+M_{22}\pm((M_{11}-M_{22})^2+4M_{12}^2)^
{1/2}\right)\numeqn$$
where by convention we take $m_{H_0}>m_{h_0}$.

The time-independent field equations are
$$\eqalign{D_j G^a_{ij}=&-\half ig\left(\Phi_\oa^\dagger\ \tau^a D_i
\Phi_\oa-(D_i\Phi_\oa)^\dagger\tau^a\Phi_\oa\right)\cr\partial_jF_{ij}=&-
\half ig'\left(\Phi_\oa^\dagger D_i\Phi_\oa-(D_i\Phi_\oa)^\dagger\Phi_\oa
\right)\cr
D_iD_i\Phi_1=&\lambda_1(\left\vert\Phi_1\right\vert^2-v_1^2)\Phi_1+\lambda_3(
\left\vert\Phi_1\right\vert^2+\left\vert\Phi_2\right\vert^2-v_1^2-v_2^2)\Phi_1
\cr&+\lambda_4(\left\vert\Phi_2\right\vert^2\Phi_1-(\Phi_2^\dagger\Phi_1)
\Phi_2)+\lambda_5(\Phi_2^\dagger\Phi_1-v_1v_2)\Phi_2\cr
D_iD_i\Phi_2=&\lambda_2(\left\vert\Phi_2\right\vert^2-v_1^2)\Phi_2+\lambda_3(
\left\vert\Phi_1\right\vert^2+\left\vert\Phi_2\right\vert^2-v_1^2-v_2^2)
\Phi_2\cr&+\lambda_4(\left\vert\Phi_1\right\vert^2\Phi_2-(\Phi_1^\dagger\Phi_2)
\Phi_1)+\lambda_5(\Phi_1^{\dagger}\Phi_2-v_1v_2)\Phi_1\cr}\numeqn$$
In addition, we recall the usual mixing formulae:
$$Z_i\equiv\cos\Theta_W W^3_i-\sin\Theta_W B_i \ ,\qquad
A_i\equiv\sin\Theta_W W^3_i+\cos\Theta_W B_i\ ,\numeqn$$

The vortex solution extremising ({\eqnint}) is given by :
$$\eqalign{&\vec W^1=0=\vec W^2=\vec A,\qquad\vec Z=-{Z(r)\over r}{\hat e}_
\theta\cr&\Phi_1=f_1(r)e^{im_1\theta}\pmatrix{0\cr1\cr},\qquad\Phi_2 = f_2(r)
e^{im_2\theta }\pmatrix{0\cr 1\cr},\cr}\numeqn$$
\rememeqn\eqnvor
where the coordinates $r$ and $\theta$ are polar coordinates
in the $xy-$plane and $m_1$ and $m_2$ are integers.
On substitution of ({\eqnvor}) into the equations of motion they reduce to
$$\eqalign{f_1''+{{f_1'}\over r}&-\left(m_1-{\alpha\over 2}Z\right)^2{f_1
\over{r^2}}-\left(\lambda_1+\lambda_3\right)\left(f_1^2-v_1^2\right)f_1\cr&-
\lambda_3\left (f_2^2 -v_2^2\right)f_1-\lambda_5\left(f_1f_2-v_1v_2\right)f_2
= 0\cr}\numeqn$$
\rememeqn\eqnmo
$$\eqalign{f_2''+{{f_2'}\over r}&-\left(m_2-{\alpha\over2}Z\right)^2{f_2\over
{r^2}}-\left(\lambda_2+\lambda_3\right)\left(f_2^2-v_2^2\right)f_2\cr&-
\lambda_3\left(f_1^2-v_1^2\right)f_2-\lambda_5\left(f_1f_2-v_1v_2\right)f_1
=0\cr}\numeqn$$
\rememeqn\eqnmt
$$Z''-{{Z'}\over r}+\alpha\left(\left(m_1-{\alpha\over2}Z\right)f_1^2+\left(
m_2-{\alpha\over2}Z\right)f_2^2\right)=0\numeqn$$
\rememeqn\eqnmz
where primes denote differentiation with respect to $r$. It is clear from the
equations that we must have $m_1=m_2$. This can also be seen by examining the
energy functional and noting that to prevent a logarithmic divergence in the
energy integral at infinity, we require that the angular dependence of
the two Higgs doublets must be identical asymptotically. We shall restrict
ourselves to the case $m_1=m_2=1$. The equations are then solved together with
the boundary conditions:
$$f_1(0)=f_2(0)=Z(0)=0,\qquad{f_1(\infty)\over v_1}={f_2(\infty)\over v_2}
={\alpha\over2}Z(\infty)=1\numeqn$$
Asymptotically, for large $r$, writing $f_1=v_1+\delta_1, f_2=v_2+\delta_2$
and $Z={2\over\alpha}+r\delta_3$
and linearising the equations of motion when $m_{H_0},m_{h_0}<m_Z$, for example
when $\lambda_i/\alpha^2$ are small, we
obtain the following modified Bessel equations
$$\pmatrix{\delta_1''+\delta_1'/r \cr \delta_2''+\delta_2'/r \cr}=
M\pmatrix{\delta_1\cr\delta_2\cr}\numeqn$$
$${\delta_3}''+{{\delta_3}'\over r}=\left(
{1\over{r^2}}+{\alpha^2\over 2}(v_1^2+v_2^2)\right)\delta_3,\numeqn$$
where $M$ is the mass matrix given by ({\eqnmas}). So, for large $r$, we
obtain $\delta_3\propto K_1(m_Zr) \sim r^{-1/2}\exp(-m_Zr)$.
Diagonalising the mass matrix $M$ by writing
${\delta_1\choose\delta_2}=\Delta_1\vec e_1 +\Delta_2\vec e_2$, where $\vec
e_1, \vec e_2$ are the eigenvectors of $M$, we get decoupled equations for
$\Delta_1$ and $\Delta_2$, yielding $\Delta_1\propto K_0(m_{H_0}r)$ and
$\Delta_2\propto K_0(m_{h_0}r)$, demonstrating that $\delta_1$ and
$\delta_2$ are linear combinations of $r^{-1/2}\exp(-m_{H_0}r)$ and
$r^{-1/2}\exp(-m_{h_0}r)$. If $m_{H_0}$ or $m_{h_0}$ is greater than
\nextref\newperiv
$2m_Z$, the asymptotics are slightly different\ref\newperiv, with $\delta_3$
source terms present in the Bessel equations for $\Delta_1,\Delta_2$
which cannot be neglected.

We note that for the particular value of the parameters
$$\tan\beta\equiv{v_2\over v_1}=\left({\lambda_1-\lambda_5\over
\lambda_2-\lambda_5}\right)^{\half},\numeqn$$
\rememeqn\eqnpar
setting $f=f_1/v_1=f_2/v_2$
causes equations ({\eqnmo}) and ({\eqnmt}) to reduce to the same equation:
$$f''+{f'\over r}-\left(1-{\alpha\over 2}Z\right)^2{f\over r^2}-\left(
(\lambda_1+\lambda_3)v_1^2+(\lambda_3+\lambda_5)v_2^2\right)(f^2-1)f
=0.\numeqn$$
In this special case, ${\delta_1\choose\delta_2}=\Delta_1{1\choose{\tan\beta}}+
\Delta_2{{\tan\beta}\choose{-1}}$. We know that $f_1/v_1=f_2/v_2$, and hence
$\delta_2/\delta_1=\tan\beta$, yielding $\Delta_2\equiv0$. This shows
that, despite the exponential dependence in $\Delta_2$ decaying more slowly
than
that in $\Delta_1$, we cannot just ignore the more rapidly decaying $\Delta_1$
when studying the asymptotic behaviour of $\delta_1$\ and $\delta_2$.

We shall examine this case in detail later since it leads to a considerable
reduction in the complexity of the stability problem.

\newsec The Harmonic Decomposition of the Perturbations

 The vortex solution given in ({\eqnvor}) is not topologically stable. The
Higgs vacuum is $S_3\times S_3$ and the first homotopy group of this is
trivial.
We are investigating the metastability of the vortex solution i.e. whether it
is a {\it local} maximum or minimum in configuration space. We consider
infinitesimal perturbations of the vortex configuration and ask if the
quadratic variation in the energy is positive or negative. We write
$$\Phi_1=\pmatrix{\ourchi_1\cr f_1(r)e^{i\theta}+\delta\phi_1\cr},\ \Phi_2=
\pmatrix{\ourchi_2\cr f_2(r)e^{i\theta} + \delta\phi_2\cr}$$
$$\eqalign{&{\vec Z}=-{1\over r}Z(r){\hat e}_{\theta}+ \delta {\vec Z},\ \ {
\vec A}=\vec a,\cr&{\vec W}^{\oa}\tau^{\oa}={\vec W}_+\tau^-+{\vec W}_-\tau^+,
\cr}\numeqn$$
where we find it convenient to use the complex notation
$$W_{\pm}={\textstyle 1\over\sqrt2}\left(W_1\pm iW_2\right),\ \tau^{\pm}=
{\textstyle 1\over\sqrt2}\left(\tau^1\pm i\tau^2\right).\numeqn$$
The perturbations can depend on the $z-$coordinate and the $z-$components
of the vector fields can be non-zero also. However, since the vortex
solution has translational invariance along the $z-$direction, it is
easy to see that the $z-$dependence in the perturbations can be ignored
and the $z-$components of the gauge fields can be set to zero.
This follows from ({\eqnint}) where
the relevant $z-$dependent terms in the integrand are:
$$\half G_{i3}^a G_{i3}^a+\fourth F_{i3}F_{i3}+(D_3\Phi_1)^{\dagger}(D_3\Phi_1)
+(D_3\Phi_2)^{\dagger}(D_3\Phi_2)\numeqn$$
This contribution to the energy is strictly non-negative and is made to vanish
by setting the $z-$components of the gauge fields to zero and also considering
the perturbations to be independent of the $z-$coordinate. For this reason,
we shall drop all reference to the $z-$coordinate
in the calculations below and it will be understood that the energy is
actually the energy {\it per unit length} of the string.

The only contribution of the photon field is the second term of the
integrand in ({\eqnint}) and this gives
$$\half ({\vec\nabla}\times{\vec a})^2$$
which is positive and hence we set ${\vec a}=0$.

The quadratic change in the energy may then be written as
$$\delta E=\int d^2x\pmatrix{\ourchi_{\oa}&\phi_{\oa},&\delta
{\vec Z},&\vec W_-\cr}^*{\bf O}\pmatrix{\ourchi_{\oa}\cr\phi_{\oa}\cr
{\delta\vec Z}\cr\vec W_-\cr},\numeqn$$
where {\bf O}, the fluctuation operator, is a second order matrix differential
operator. The perturbations will then decompose into channels labelled by the
eigenvalues of the operators generating the continuous symmetries of {\bf O}
and by the eigenvalues of the discrete symmetries of {\bf O}. These symmetries
are the subset of the symmetries of
the energy density which are also symmetries of the vortex configuration
({\eqnvor}).

It has a $Z_2$ symmetry under the action of the large $U(2)$ gauge
transformation given by
$${\overline U}=\pmatrix{-1&0\cr 0&1\cr}.\numeqn$$
The perturbations of the string itself,
$\delta\phi_{\oa}$ and $\delta{\vec Z}$ , lie in the ${\overline U}=1$ channel
and $\ourchi_{\oa}$ and ${\vec W}^{\oa}\tau^{\oa}$ are in ${\overline U}=-1$.
Hence these sets of perturbations decouple and we can write
$$\delta E=\delta E_1(\delta\phi_{\oa},\delta{\vec Z})+\delta
E_2(\ourchi_{\oa},W_-\tau_+).\numeqn$$
The first term is positive since it corresponds to the stability problem of the
vortex in the abelian Higgs case which is topologically stable as discussed
\nextref\la
in reference [\la]. Hence we set $\delta\phi_{\oa}=0,\ \delta{\vec Z}=0.$

The configuration is axially symmetric i.e. it is invariant under the action of
the symmetry operator generated by the generalised angular momentum operator
$$K_z\ =\ L_z\ +\ S_z\ + \ I_z \numeqn $$
where $L_z$ and $S_z$ are the usual
orbital and spin pieces respectively of the spatial angular momentum operator
given explicitly by
$$L_z=-i{\partial\over \partial\theta}{\bf 1}\ ,\ (S_z{\vec
b})_j=-i\epsilon_{3jk}{\vec b_k}{\bf 1}\ ,\numeqn$$
where ${\bf 1}$ is the unit matrix and $\vec b$ is any vector field (note that
$S_z$ annihilates the Higgs doublets). $I_z$ is
composed of a $U(1)$ generator, $Y$ and an $SU(2)$ generator, $T^3$
$$ I_z= -\half( Y-T^3)\ . \numeqn $$
We therefore perform a harmonic expansion of the perturbations,
$$\eqalign{\ourchi_{\oa}&=\sum_m\ourchi_{\oa}^m(r)e^{im\theta}\cr
{\vec W}_-\tau^+&=\sum_n\left(-iF_n(r){\hat e}_r+{\xi_n(r)\over r}
{\hat e}_{\theta}\right)e^{i(n-1)\theta}\tau^+\,\cr}\numeqn$$
where m and n take values from $-\infty$ to $+\infty$. It is clear that this is
a decomposition into angular momentum modes, since
$$K_z(e^{im\theta})=me^{im\theta}\ ,\ \
K_z(e^{i(n-1)\theta}\tau^+)=ne^{i(n-1)\theta}\ .\numeqn$$
The configuration is also CP-invariant i.e. it is invariant under the
combination of reflection in the x-axis and complex conjugation. Therefore the
channels may be labelled as CP-even and CP-odd. It is straightforward to check
that the $U(2)$ gauge transformation $U={\rm diag}[i,1]$ swaps the two
channels and hence the two instability problems are exactly equivalent. Hence
this supplies a reality condition and we may
take $\ourchi^m_{\oa}, F_n(r)$ and $\xi_n(r)$ to be real.

\newsec Analysis of the Least Stable Mode

We have shown that the relevant energy variation can be decomposed into a sum
over angular momentum modes
$$\delta E=\sum_m\delta E\left(\ourchi^m_{\oa}, F_m(r), \xi_m(r)\right).
\numeqn$$

Explicitly for the ${\rm m}^{\rm th}$ mode we split the variation in the
energy into three contributions
$$\delta E\left(\ourchi^m_{\oa}, F_m(r), \xi_m(r)\right)=E_h+E_c+E_W\numeqn$$
where $E_h$ arises solely from the perturbation in the Higgs fields, $E_W$ is
from the perturbation of the W-fields and $E_c$ is from the interaction.

Then $$\eqalign
{E_h=2\pi\int_0^{\infty}rdr\biggl[&\left\vert{\ourchi_1^m}'\right\vert^2+
\left\vert{\ourchi_2^m}'\right\vert^2+{1\over r^2}\left(m+{\alpha \over
2}\cos2\Theta_WZ\right)^2(\left\vert\ourchi_1^m\right\vert^2+\left\vert\ourchi_2
^m\right\vert^2)\cr
&+(\lambda_1+\lambda_3)(f_1^2-v_1^2)\left\vert\ourchi_1^m\right\vert^2+(
\lambda_2
+\lambda_3)(f_2^2-v_2^2)\left\vert\ourchi_2^m\right\vert^2\cr&+\lambda_3((f_1^2
-v_1 ^2)\left\vert\ourchi_2^m\right\vert^2+(f_2^2-v_2^2)\left\vert\ourchi_1^m
\right\vert^2+\lambda_4\left\vert
f_2\ourchi_1^m-f_1\ourchi_2^m\right\vert^2\cr&+2\lambda_5(f_1f_2-v_1v_2)
{\ourchi_1^m}\ourchi_2^m\biggr]\cr}
\numeqn$$
\rememeqn\eqneh
$$\eqalign{E_c=2\pi g\int_0^{\infty}rdr\biggl[&
\bigl(({f_1}'\ourchi_1^m-f_1{\ourchi_1^m}')
+({f_2}'\ourchi_2^m-f_2{\ourchi_2^m}')\bigr)F_m\cr
&\qquad\qquad-{\xi_m\over
r^2}(f_1\ourchi_1+f_2\ourchi_2)\left(m+1-\alpha\sin^2\Theta_WZ\right)
\biggr]\cr} \numeqn$$
$$E_W=\pi\int_0^{\infty}{dr\over r}\biggl[2F_m\xi_m\gamma
Z'+\left(\left(m-(1-\gamma Z)\right)F_m-\xi_m')\right)^2+{g^2(f_1^2+f_2^2)
\over 2}(\xi_m^2+r^2F_m^2)\biggr]\numeqn$$
\rememeqn\eqnew
An instability is most likely to develop in the core of the string, away from
the vacuum manifold. Considering the Higgs perturbations, if we examine
${\rm E}_h$ ({\eqneh}) we see that a Higgs condensate in the core could lead
to a decrease in the potential energy. The Higgs perturbations can only be
non-vanishing in the core for m=0, else the field is singular there.

If we now consider ${\rm E}_W$ ({\eqnew}) we see that the only term that can
make a negative contribution is the first term of the integrand. The physical
\nextref\perkins
interpretation of this instability has been elucidated by Perkins\ref{\perkins}
: the
magnetic field $\vec B=\vec\nabla\times\vec Z$ in the core of the string
couples to the anomalous magnetic moment of the W-bosons and they acquire a
negative effective mass. The potential instability is thus a W-condensate
in the core. If we perform the Taylor expansion around the origin
$$F_m=a_{m0}+a_{m2}r^2+\cdots\ ,\ \xi_m=rb_{m0}+r^3b_{m2}+\cdots\ ,\numeqn$$
then we find that
$${\vec W}_-\tau^+=\sum_n\pmatrix{-i(a_{m0}\cos\theta-ib_{m0}\sin\theta)\cr-i
a_{m0}\sin\theta+b_{m0}\cos\theta\cr} e^{i(m-1)\theta} \tau^++O(r^2)\
.\numeqn$$
Then the requirement that the fields be non-singular at the origin gives that
they can only be non-vanishing there for m$=0,2$. First considering the case
$m=2$, we then require ${F_2}={\xi_2}/r$, that is they have the
same sign and the contribution is therefore positive since Z$'$ is positive in
the core. However, for the case $m=0$ the condition is that $F_0=-{\xi_0}/r$
and
the contribution is then negative.

We conclude that the least stable mode is the $m=0$ mode. This confirms the
intuitive idea that we expect the least stable mode to have $K_z=0$ since as
$\bigl\vert K_z\bigr\vert$ increases one gets an increasing centrifugal
barrier. We henceforth restrict ourselves to the case m$=0$. If we examine
(\eqneh--\eqnew) we note that there are no $F'$ terms
(we drop the m-script from now on). We therefore complete the squares and
obtain
$$\eqalign{\delta E=\pi\int_0^{\infty}{dr\over r}&\biggl[{\xi'}^2+\half
g^2(f_1^2+f_2^2)\xi^2-{Q^2\over P_+}-2g(1-\alpha\sin^2\Theta_WZ)\xi(f_1
\ourchi_1+f_2\ourchi_2)\cr&+2r^2({{\ourchi_1}'}^2+{{\ourchi_2}'}^2)+\half(
\alpha\cos2\Theta_WZ)^2(\ourchi_1^2+\ourchi_2^2)\cr
&+2r^2\Bigl\{(\lambda_1+\lambda_3)(f_1^2-v_1^2)\ourchi_1^2+(\lambda_2+\lambda_3
)(f_2^2-v_2^2)\ourchi_2^2+\lambda_4(f_2\ourchi_1-f_1\ourchi_2)^2\cr
&+\lambda_3((f_1^2-v_1^2)\ourchi_2^2+(f_2^2-v_2^2)\ourchi_1^2)+2\lambda_5(f_1
f_2-v_1v_2)\ourchi_1\ourchi_2\Bigr\}\biggr]\cr
&\kern-1.5em +T(F,\ourchi_1,\ourchi_2,\xi)\cr}\numeqn$$ where
$$T(F,\ourchi_1,\ourchi_2,\xi)=\pi\int_0^\infty {dr\over r}\left(
   \sqrt{P_+}F+{Q\over{\sqrt{P_+}}}\right)^2,\numeqn$$
and
$$Q=\xi'(1-\gamma Z)+\gamma Z'\xi-gr^2\left(f_1{\ourchi_1}'-{f_1}'\ourchi_1+f_2
{\ourchi_2}'-{f_2}'\ourchi_2\right)\ ,\numeqn$$
and
$$P_+=(1-\gamma Z)^2+\half g^2r^2(f_1^2+f_2^2)\ .\numeqn$$
The contribution from  $ T(F,\ourchi_1,\ourchi_2,\xi)$ is positive and we have
an algebraic equation for $F$, namely
$$F=-{Q\over P_+}\ ,\numeqn$$
so that the contribution is made to vanish.
We are now left with a problem in three variables and
we express the change in the energy in the form
$$\delta E[\ourchi_1,\ourchi_2,\xi]=2\pi\int dr\,r(\ourchi_1,\ourchi_2,\xi)
{\bf{\bar O}}\pmatrix{\ourchi_1\cr\ourchi_2\cr\xi\cr},\numeqn$$
\rememeqn\eqne
where ${\bf {\bar O}}$ is a $3\times 3$ matrix differential operator.

We have not completely fixed the gauge, and hence there is a gauge zero
mode which we must extract. Perturbations of the form
$$\delta\Phi=ig\psi\Phi_0,{\ {\delta W}_i^a\tau^a=-iD_{0i}\psi},\numeqn$$
where $\psi$ is a real $L(SU(2))$
valued function and the $0$ subscript denotes the unperturbed
fields, give an infinitesimal gauge transformation of the
vortex solution ({\eqnvor}). If we now require that the gauge perturbations lie
in $( \ourchi_1,\ourchi_2 , \xi )$ then we can only have
$$\psi=s(r)\pmatrix{0&ie^{-i\theta}\cr -ie^{i\theta}&0\cr}\numeqn$$
where $s(r)$ is any smooth function. This means that
perturbations given by
$$\pmatrix{\ourchi_1\cr\ourchi_2\cr\xi}\ =\ s(r)
\pmatrix{-gf_1\cr-gf_2\cr 2(1-\gamma Z)\cr}\numeqn$$
are pure gauge perturbations and are annihilated by ${\bf {\bar O}}$. Hence we
need a basis for our physical perturbations orthogonal to this. We choose
$$\eqalign{\pmatrix{\ourchi_1\cr\ourchi_2\cr v\xi\cr}=&
\pmatrix{gf_1 /2\cr gf_2 /2\cr -v(1-\gamma Z)\cr}s_0
(f_1^2+f_2^2)^{-1/2}+ \pmatrix{f_2\cr -f_1\cr 0\cr}s_1
(f_1^2+f_2^2)^{-1/2}\cr
&+\pmatrix{f_1(1-\gamma Z)\cr f_2(1-\gamma Z)\cr {g\over {2v}}
(f_1^2+f_2^2)\cr}{s_2\over P}(f_1^2+f_2^2)^{-1/2},}\numeqn$$
\rememeqn\eqnw
where $v$ is a mass-scale so that the basis
vectors have consistent dimensions and
$$P=(1-\gamma Z)^2+{g^2\over{4v^2}}(f_1^2+f_2^2).\numeqn$$
Before substituting (\eqnw) into (\eqne), to make the dependence on the
parameters $t=\tan\beta=v_2/v_1$, $\sin^2\Theta_W$, $\beta_i=\lambda_i/
\alpha^2$ explicit we rescale the fields as follows
$$R=vr,\ \ v_1F_1=f_1,\ \ v_2F_2=f_2,\ \ \bz=\half\alpha Z\ ,\numeqn$$
\rememeqn\eqnrescale
where we now set
$$v=\alpha(v_1v_2)^{\half}\ .\numeqn$$
$P_+$ and $P$ are now re-expressed as
$$\eqalign{P_+&=(1-\gz)^2+\half R^2\csqw F^2\cr
P=&(1-\gz)^2+{\textstyle {1\over 4}}\csqw F^2\cr}\numeqn$$
where
$$F^2={1\over t}F_1^2+tF_2^2.\numeqn$$
On substitution of ({\eqnw}) into ({\eqne}) it was a good check on our algebra
that the $s_0$ terms vanished. After a lot of algebra we obtain
$$\delta E=\pi\int_{0}^{\infty}dR\left({{\underline s}'}^TM{\underline s}'
+{\underline s}^TN{\underline s}'+{\underline s}^T
S{\underline s}\right)\numeqn$$
\rememeqn\eqnfin
where $'$ now denotes differentiation with respect to R, $s^T=(s_1,s_2)$, and
$$M=\pmatrix{M_{11}&0\cr0&M_{22}\cr},\quad N=\pmatrix{0&N_{12}\cr0&0\cr}
,\quad S=\pmatrix{S_{11}&S_{12}\cr S_{12}&S_{22}\cr}.\numeqn$$
We give the coefficients explicitly in the appendix. Note that M is diagonal
as one would expect from choosing orthogonal coordinates. Also, it is of
interest to note that $N_{12}$ and $S_{12}$, the coefficients giving the
coupling between $s_1$ and $s_2$, are both proportional to $(F_1F'_2-F_2F'_1)$
so that for the special choice of parameters ({\eqnpar}), which gives
$F_1=F_2$, $N_{12}$ and $S_{12}$ are zero and the equations decouple.

We are only assured of a complete set of eigenvectors and hence of
obtaining all possible eigenvalues when the problem is posed in a self-adjoint
form. The problem as it is posed in ({\eqnfin}) is not in self-adjoint form due
to the non-vanishing coefficient $N_{12}$. Hence we must perform a further
change of variables
$$\pmatrix{u_1\cr u_2\cr}=\pmatrix{s_1\cr s_2+\lambda
s_1\cr}\ .\numeqn$$
If we then choose
$$\lambda'={N_{12}\over
2M_{22}}=2(1-\gz){(F_1 F_2'-F_2 F_1')\over F^2},$$
which we
solve together with the boundary condition $\lambda(0)=0$ then the
problem ({\eqnfin}) is now expressed in self-adjoint form:
$$\delta E=\pi\int_{0}^{\infty}dR\left({\underline u}'^T{\bar M}{\underline
 u}'+{\underline u}^T{\bar S}{\underline u}\right)\ ,\numeqn$$
where trivially ${\bar M}$ and ${\bar S}$ are symmetric. Now to find the
coupled Schr\"odinger equation we extremise $\delta E$ subject to fixing the
weighted norm of ${\underline u}$ via
$$2\pi\int_0^{\infty}\!RdR\,{\underline u}^T{\bar W}{\underline u}=1,$$
where
$${\bar W}=\pmatrix{1+\lambda^2/P_+&-\lambda/P_+\cr-\lambda/P_+&1/P_+\cr}.
\numeqn$$
We then introduce the Lagrange multiplier $\mu$ and extremise
$$\pi\int_0^{\infty}dR\left({\underline u}'^T{\bar M}{\underline u}'+
{\underline u}^T{\bar S}{\underline u}-2\mu R{\underline u}^T{\bar W}
{\underline u}\right),\numeqn$$
which gives the coupled Schr\"odinger equations
$$-\left({\bar M}{\underline u}'\right)'+{\bar S}{\underline u}=2\mu R{\bar W}
{\underline u}.\numeqn$$
This is equivalent to extremising
$$\pi\int_{0}^{\infty}dR\left({{\underline s}'}^TM{\underline s}'
+{\underline s}^TN{\underline s}'+{\underline s}^T S{\underline s}
-2\mu R{\underline s}^T W{\underline s}\right)\numeqn$$
where
$$W=\pmatrix{1&0\cr0&1/P_+\cr},\numeqn$$
which gives the coupled equations
$$\eqalign{&2(M_{11}s'_1)'=N_{12}s'_2+2(S_{11}-2\mu R)s_1+2S_{12}s_2\cr
&2(M_{22}s'_2)'+(N_{12}s_1)'=2(S_{22}-2\mu R/P_+)s_2+2S_{12}s_1.\cr}\numeqn$$
\rememeqn\eqnprev
The boundary conditions on $s_1$ and $s_2$ are taken to be
$$s_1^2(0)+s_2^2(0)=1,\ \ s_1(\infty)=s_2(\infty)=0,\numeqn$$
which is permissible due to the linearity of the problem,
and the ratio $s_1(0)/s_2(0)$ will vary according to the eigenvalue $\mu$.
It then follows that the stability of the string is dependent on $\mu$, since
for the solutions of ({\eqnprev}),
$$\delta E=\mu\int_{0}^{\infty}\!RdR\,{\underline s}^TW{\underline s},\numeqn$$
and since $W$ and $\bar W$ are strictly positive definite then if $\mu$ is
negative the string is unstable, and if $\mu$ is positive the string is stable.

\newsec Numerical Studies

On making the change of variables (\eqnrescale), equations (\eqnmo--\eqnmz)
become
$$\eqalign{F_1''+{{F_1'}\over R}&-\left(1-\bz\right)^2{F_1\over{R^2}}-
{1\over t}\left(\beta_1+\beta_3\right)\left(F_1^2-1\right)F_1\cr&-
t\beta_3\left(F_2^2-1\right)F_1-t\beta_5\left(F_1F_2-1\right)F_2=0\cr}\numeqn$$
$$\eqalign{F_2''+{{F_2'}\over R}&-\left(1-\bz\right)^2{F_2\over{R^2}}-t\left(
\beta_2+\beta_3\right)\left(F_2^2-1\right)F_2\cr&-{1\over t}\beta_3\left(F_1^2-
1\right)F_2-{1\over t}\beta_5\left(F_1F_2-1\right)F_1=0\cr}\numeqn$$
$$\bz''-{{\bz'}\over R}+\12\left(1-\bz\right)F^2=0\numeqn$$
which are solved together with the boundary conditions
$$F_1(0)=F_2(0)=\bz(0)=0,\qquad F_1(\infty)=F_2(\infty)=\bz(\infty)=1\numeqn$$
by relaxation.
To determine the minimum eigenvalue $\mu$, the equations (\eqnprev) are then
solved by relaxation on the same lattice. Since the couplings between $s_1$ and
$s_2$ are small, the spectrum of eigenvalues can be split into two families;
one in which $\underline s$ is mainly $s_1$, and one in which $\underline s$
is mainly $s_2$. These families are studied separately by solving equations
(\eqnprev) in two separate cases, namely by imposing either $s_1(0)=1$ or
$s_2(0)=1$, to yield two minimal eigenvalues which we will call $\mu_1$ and
$\mu_2$ respectively.

If we set $\tan\beta=1$, $\beta_1=\beta_2$, $\beta_3=\beta_4=\beta_5=0$, then
condition (\eqnpar) is satisfied and the $s_1$ and $s_2$ equations decouple.
Writing $\Phi_1=\Phi_2=\Phi/\sqrt2$ causes our energy functional
to reduces to that of the standard electroweak model, provided we set
$\beta_1=\beta_2=\half\beta$, where $\sqrt\beta$ is the ratio of the Higgs
and Z masses. In this special case, we found the same region of stability
for $s_2$ perturbations as that found in references [\vachaspatiE,\james],
which
provided a useful check on the reliability of our numerics.

It was found that $\mu_1$ is typically positive, giving no unstable modes made
principally of $s_1$ perturbations. Since $\mu_2$ depends only very weakly on
$\beta_4$, a negative $\mu_1$ can easily be made positive by increasing the
value of $\beta_4$ slightly without seriously affecting $\mu_2$. In the search
for parameters which give a stable vortex solution, the main difficulty is in
making $\mu_2$ positive.
Over a wide range of parameters, it was found that the crucial parameter in
determining $\mu_2$ is
$${{m_{H_0}^2}\over{m_Z^2}}={{2t}\over{1+t^2}}\left\{
c_1+c_2+\left(t+{1\over t}\right)c_5+\left[\left(c_1-c_2+\left(t-{1\over t}
\right)c_5\right)^2+c_3^2\right]^{1/2}\right\},\numeqn$$
\rememeqn\eqnratio
where $c_1=(\beta_1+\beta_3)/t$, $c_2=t(\beta_2+\beta_3)$,
$c_3=2\beta_3+\beta_5$, $c_5=\beta_5/2$. Moving around the parameter space on
curves of fixed $m_{H_0}/m_Z$ typically results in little variation in $\mu_2$.
Hence, to analyse the stability properties, we can restrict ourselves to the
case where our energy functional reduces to that of the standard electroweak
theory, thus reproducing the stability properties of
references [\vachaspatiE,\james].
Numerical studies demonstrate that by tweaking the parameters in the model for
fixed mass ratio (\eqnratio) $\mu_2$ can be slightly increased from the value
in the electroweak model, but this increase is small. Some sample results are
displayed in table 1.
Hence, since the region of stability for strings in the standard electroweak
model is a long way from experimentally acceptable values of the Weinberg angle
and Higgs to Z mass ratio, the increase in the region of stability obtained by
adding a second Higgs doublet is insufficient to lead to stable strings for
realistic parameters.
\def\tablerule{\noalign{\hrule}}
\newbox\mystrutbox
\setbox\mystrutbox=\hbox{\vrule height11.5pt depth5.5pt width0pt}
\def\mystrut{\relax\ifmmode\copy\mystrutbox\else\unhcopy\mystrutbox\fi}
\def\results{\vbox{\tabskip=0pt\offinterlineskip
\halign to 5.4true in {\mystrut##&\vrule##\tabskip=1em plus 2em&
\hfil##\hfil&\vrule##&
\hfil##\hfil&\vrule##&
\hfil##\hfil&\vrule##&
\hfil##\hfil&\vrule##&
\hfil##\hfil&\vrule##&
\hfil##\hfil&\vrule##&
##\hfil&\vrule##&
##\hfil&\vrule##
\tabskip=0pt\cr\tablerule
&&\omit\hidewidth$\sin^2\Theta_W$\hidewidth&&
  \omit\hidewidth$\tan\beta$\hidewidth&&
  \omit\hidewidth$\beta_1$\hidewidth&&
  \omit\hidewidth$\beta_2$\hidewidth&&
  \omit\hidewidth$\beta_3$\hidewidth&&
  \omit\hidewidth$\beta_5$\hidewidth&&
  \omit\hidewidth${\cal R}$\hidewidth&&
  \omit\hidewidth$\mu_2$\hidewidth&\cr\tablerule
&&0.95&&1.0&&0.8&&0.8&&0.0&&0.0&&1.6&&--0.12${}^*$&\cr\tablerule
&&0.95&&1.0&&0.2&&0.2&&0.2&&0.2&&1.6&&--0.12&\cr\tablerule
&&0.95&&1.7&&1.17&&0.8&&--0.4&&0.5&&1.6&&--0.10&\cr\tablerule
&&0.95&&2.8&&0.44&&0.35&&--0.12&&0.6&&1.6&&--0.13&\cr\tablerule
&&0.95&&1.0&&1.6&&0.0&&0.0&&0.0&&1.6&&--0.28&\cr\tablerule
&&0.6&&1.0&&0.2&&0.2&&0.0&&0.0&&0.4&&--0.57${}^*$&\cr\tablerule
&&0.6&&1.0&&0.05&&0.05&&0.05&&0.05&&0.4&&--0.57&\cr\tablerule
&&0.6&&0.8&&0.04&&0.17&&0.02&&0.05&&0.4&&--0.52&\cr\tablerule
&&0.6&&1.0&&0.4&&0.0&&0.0&&0.0&&0.4&&--0.52&\cr\tablerule
&&0.6&&0.2&&0.208&&0.0&&0.0&&0.0&&0.4&&--1.86&\cr\tablerule
&&0.6&&2.0&&1.0&&0.0&&0.0&&0.0&&0.4&&--0.45&\cr\tablerule
}}}
\smallskip
\centerline{Table 1}
\medskip\centerline{\results}
\smallskip\noindent
In table 1, ${\cal R}=m_{H_0}^2/m_Z^2$, and $\beta_4=0$ in all cases. In the
starred cases, our model reduces to the one doublet electroweak theory.

\newsec Conclusion

In section 2 we demonstrated that there are vortex solutions in the
CP-invariant two Higgs doublet electroweak theory throughout the allowable
parameter range. The `width' of such vortices depends on the masses of the two
CP-even neutral Higgs particles $H^0$ and $h^0$, and on the $Z$-mass. In
section
3 a general perturbation of such a vortex was decomposed into angular momentum
eigenstates, and it was seen that the perturbations decouple into two parts --
one the same as in the two singlet abelian Higgs model, and a second part
arising from nonzero $\vec W$ and Higgs perturbations orthogonal to the
original profiles. The CP-invariance of the model was used to decouple the real
and imaginary parts of the perturbations. In section 4, the least stable mode
in our expansion of the perturbations was picked out after linearisation, and
a change of coordinates was performed to make explicit the component of this
mode which was an infinitesimal gauge transformation. The change in energy was
then expressed in self-adjoint form, to be analysed as a Sturm-Liouville
eigenvalue problem. The numerical study of the perturbations was discussed in
section 5, and it was discovered that the most important determining factor in
the sign of the energy associated with the perturbation was the ratio of the
masses of the $H^0$ and $Z$ particles. The region of stability was only
slightly larger than that for vortices in the one doublet electroweak model,
and we conclude that for realistic values of the Weinberg angle and Higgs mass,
vortex solutions in the two doublet electroweak model are unstable.

\beginsec Acknowledgements

We thank Anne-Christine Davis for her advice and for encouraging us in doing
this problem. Both authors acknowledge a SERC research student grant.

\beginsec Appendix

In section 4. we showed that the stability problem is reduced to extremising
$$\pi\int_{0}^{\infty}dR\left({{\underline s}'}^TM{\underline s}'
+{\underline s}^TN{\underline s}'+{\underline s}^TS{\underline s}
-2\mu R{\underline s}^TW{\underline s}\right)\numeqn$$
where $'$ denotes differentiation with respect to $R$.
$$M_{11}=P_+M_{22}=2R$$
$$N_{12}=8R(1-\gz){(F_1F'_2-F'_1F_2)\over{P_+F^2}}$$
$$\eqalign{S_{11}=&2R\left[{{F_1F'_2-F'_1F_2}\over{F^2}}\right]^2
\left(1-2\csqw R^2F^2P_+^{-1}\right)+{2\over R}\csqw\bz^2+2R\beta_4F^2\cr
&+{2R\over F^2}\biggl[(\beta_1+\beta_3)(F_1^2-1)F_2^2
+(\beta_2+\beta_3)(F_2^2-1)F_1^2\cr
&\qquad\qquad+\beta_3\left({1\over t^2}(F_1^2-1)F_1^2+t^2(F_2^2-1)F_2^2)
\right)-2\beta_5(F_1F_2-1)F_1F_2\biggr]\cr}$$
$$S_{12}=4R{(F_1F'_2-F'_1F_2)\over{P_+F^2}}\left[\gz'-(1-\gz){F'\over F}
\right]$$
$$S_{22}=A_1+A_2-\12 A_3'$$
where
$$\eqalign{A_1=&{1\over8}\cos^4\Theta_W{R\over{P^2P_+}}\left(FF'\right)^2-\12
\csqw{P'\over{RP^3}}FF'+\fourth\csqw{{P'^2}\over{RP^4}}F^2\cr
&+{1\over8}\cos^4\Theta_W{F^4\over{RP^2}}+2\cos^2 2\Theta_W(1-\gz)^2
{\bz^2\over{RP^2}}\cr
&-\csqw(1-\gz)(1-2\sin^2\Theta_W\bz){F^2\over{RP^2}}\cr
&+{{2R}\over{P^2}}\left[\gz'+(1-\gz)P'/P \right]^2\cr
&+{{2R}\over{P^2}}(1-\gz)^2\left[\left({{F_1F_2'-F_1'F_2}\over{F^2}}\right)^2
-\12\csqw{{F'^2}\over{P_+}}(1+R^2)\right]\cr
&-\csqw{F^2\over{RP^2P_+}}\left[(\half+R^2)\gz'-(\half-R^2)(1-\gz)P'/P
\right]^2\cr
&-2\csqw{{FF'}\over{RP^2P_+}}(\half+R^2)\times\biggl[(\half+R^2)\gz'(1-\gz)\cr
&\qquad\qquad\qquad-(\half-R^2)(1-\gz)^2P'/P\biggr]\cr}$$
$$\eqalign{A_2=&{{2R}\over{P^2F^2}}(1-\gz)^2\biggl[(\beta_1+\beta_3)(F_1^2-1)
{F_1^2\over t^2}+(\beta_2+\beta_3)(F_2^2-1)F_2^2t^2\cr
&\qquad\qquad+\beta_3\left((F_1^2-1)F_2^2+(F_2^2-1)F_1^2\right)+
2\beta_5(F_1F_2-1)F_1F_2\biggl]\cr}$$
$$\eqalign{A_3=&{1\over{RP^2P_+}}\left[\csqw
R^2FF'-2(1-\gz)\gz'\right]\times\cr
&\qquad\left[2R^2(1-\gz)^2+\fourth\csqw F^2\right]-4R{P'\over{PP_+}}\cr}$$

\beginsec References

\frenchspacing

\REF{\vachaspatiA}{T. Vachaspati, M. Barriola}{Phys. Rev.
Lett.}{69}{1992}{1867}

\item{}{\reffont M. Barriola, T Vachaspati, M. Bucher}\ preprint TUTP--93--7

\REF{\vachaspatiB}{T. Vachaspati, A. Ach\'ucarro}{Phys. Rev.}{D44}{1991}{3067}

\REF{\nielsen}{H. B. Nielsen, P. Olesen}{Nucl. Phys.}{B61}{1973}{45}

\REF{\hindmarsh}{M. Hindmarsh}{Phys. Rev. Lett.}{68}{1992}{1263}

\REF{\vachaspatiC}{A. Ach\'ucarro, K. Kuijken, L. Perivolaropoulos, T.
Vachaspati,}{Nucl. Phys.}{B388}{1992}{435}

\REF{\vachaspatiD}{T. Vachaspati}{Phys. Rev. Lett.}{68}{1992}{1977}

\REF{\nambu}{Y. Nambu}{Nucl. Phys.}{B130}{1977}{505}

\REF{\vachaspatiE}{M. James, T. Vachaspati, L. Perivolaropoulos}{Phys.
Rev.}{D46}{1992}{5232}

\REF{\james}{M. James, T. Vachaspati, L. Perivolaropoulos}{Nucl.
Phys.}{B395}{1993}{534}

\REF{\vachaspatiF}{R. Holman, S. Hsu, T. Vachaspati, R. Watkins}{Phys. Rev.}
{D46}{1993}{5352}

\REF{\davis}{R. Brandenberger, A-C. Davis}{Phys. Lett.}{B308}{1993}{79}

\item{[\vachaspatiG]}{\reffont T. Vachaspati, R. Watkins}\ preprint
hep-ph/921184

\item{[\higgs]}{\reffont J.F. Gunion, H.E. Haber, G.L. Kane, S. Dawson}\ `The
Higgs Hunter's Guide' Addison Wesley (1989)

\item{[\newperiv]}{\reffont L. Perivolaropoulos}\ preprint (1993)

\item{[\la]}{\reffont H.S. La}\ preprint hep-ph/9302220

\REF{\perkins}{W.B. Perkins}{Phys. Rev.}{D47}{1993}{5224}

\bye